\documentclass[useAMS,usenatbib]{mn2e}

\usepackage{graphicx}
\newcommand{\mas}{mas}
\newcommand{\may}{mas\,yr$^{-1}$}

\newcommand{\kms}{km\,s$^{-1}$}
\newcommand{\vs}{$v \sin i$}
\newcommand{\teff}{$T_{\rm eff}$}
\newcommand{\lgg}{$\log\,{g}$}

\title[Rotation and Abundances in Ap Members of NGC 6475]{Rotation and chemical abundances of Ap/Bp stars in the open cluster NGC 6475}

\author[Folsom et al.]{C.P. Folsom$^{1,2}$\thanks{E-mail: Colin.Folsom@rmc.ca}, G.A. Wade$^{1}$, S. Bagnulo$^{3}$, and J.D. Landstreet$^{4}$ \\
$^{1}$Department of Physics, Royal Military College of Canada, P.O. Box 17000, Station `Forces', Kingston, Ontario, Canada, K7K 7B4\\
$^{2}$Department of Physics, Engineering Physics \& Astronomy, Queen's University, Kingston, Ontario, Canada, K7L 3N6 \\
$^{3}$European Southern Observatory, Casilla 19001, Santiago 19, Chile\\
$^{4}$Physics \& Astronomy Department, The University of Western Ontario, London, Ontario, Canada, N6A 3K7 }

\begin{document}

\date{Received ??; Accepted ??}

\pagerange{\pageref{firstpage}--\pageref{lastpage}} \pubyear{2006}

\maketitle

\label{firstpage}

\begin{abstract}
The connection between age, rotation and chemical abundance of magnetic Ap stars is poorly understood.
Using open clusters, we are able to study samples of stars that are both co-eval and co-environmental. By determining 
rotation and chemical abundance for Ap star members of clusters with various ages, the variations of these properties 
as a function of age and environment can be derived.
All four probable Ap star members of the open cluster NGC 6475, as well as one normal late B star, 
were studied using detailed spectrum synthesis of high resolution UVES-POP spectra.  
Probable cluster membership was confirmed for all five stars, however chemical abundance anomalies only appear to be 
present in spectra of three. Projected rotational velocity and chemical abundances for 21 elements ranging from C to Eu are 
presented for the 5 stars. In the three peculiar stars we find overabundances of Si, Cr, Mn, Fe and  rare earths such as Nd, 
characteristic of Ap stars.  The set of chemically peculiar stars show fairly homogeneous abundance tables, 
however notable differences exist for a few elements.  
There also exist appreciable differences in the \vs\, and main sequence evolutionary stage of the 
chemically peculiar stars.  This may hint at the underlying processes giving rise to the observed abundance anomalies. 
With this first detailed study of chemical abundances of a complete sample of magnetic Ap/Bp stars in an open cluster, 
we have initiated an exploration of the environmental and evolutionary influence on chemical peculiarity. 
\end{abstract}


\section{Introduction}
Ap and Bp stars are intermediate-mass main sequence stars that display strong photospheric abundance anomalies.  
In particular, iron peak elements can be strongly overabundant compared to solar: Fe by +1 dex and Cr up to +3 dex.  
Si can also be overabundant by +1 dex and rare earth elements, such as Nd and Eu, can be even more 
strongly over abundant in the star's photosphere.  In addition to these strong peculiarities,
Ap and Bp stars also possess strong magnetic fields ($\sim 1$ kG) that are ordered on large scales; these are predominantly dipolar in structure. 
This is remarkable since intermediate mass stars do not possess the deep convective envelopes needed to power a 
conventional dynamo, as seen in lower mass stars.  There certainly exists a link between the strong magnetic fields and the 
observed chemical peculiarities.  In rough terms, the abundance anomalies are a result of the competing effects of radiative levitation 
and gravitational settling, producing a chemically stratified atmosphere.  Elements with a local accumulation near the surface of the star appear 
overabundant.  The magnetic field is thought to provide additional stability to the atmosphere allowing horizontal 
inhomogeneities and vertical stratification to persist.  However, this picture is still incomplete.  
The details of the stratification process and the effects of the 
magnetic field are not understood.  As well, the origin of the magnetic field remains a mystery. 
Perhaps most glaringly, the evolution of magnetic fields and chemical anomalies in Ap/Bp stars 
during the main sequence phase remains unknown.  

Significant effort has gone into examining specific Ap and Bp stars: modelling chemical abundances, magnetic 
fields, and non-uniform distributions of elements both vertically and horizontally across a star's surface. 
Despite a growing sample of well modelled Ap/Bp stars, our ability to describe general trends 
as these stars age is limited.  Determining the age of an isolated star to a high degree of accuracy 
can be difficult.  Moreover, Ap/Bp stars display abnormal photometric properties, which can 
greatly exacerbate errors in the dating of a star.  (See Bagnulo et al. 2006 for further discussion of 
these difficulties.)  Consequently, detailed analysis of Ap/Bp 
stars (often referred to as simply Ap stars) in open clusters can be of great benefit.  
The large sample of normal stars in an open cluster enables one to determine a much more 
accurate age then is possible for a lone field Ap star, which can then be applied to all stars 
in the cluster.  With accurate ages and detailed properties, a proper description of the 
evolution of Ap stars can be obtained.  A few papers have dealt with chemical abundances analysis of 
intermediate mass stars in open clusters.  Burkhart and Coupry (1998) examined the Praesepe open 
cluster and found a number of Am stars.  Varenne and Monier (1999) investigated a large sample of stars in 
the Hyades cluster and reported abundances for normal A and F stars, as well as a number of chemically peculiar stars.  
St\"utz et al. (2006) looked at 5 stars in the open cluster IC 2391 and found some evidence for chemical peculiarity in one.  
Despite this beginning, much work remains before we can properly describe the evolution of Ap stars.
This paper focuses on one cluster in particular, the 220 Myr old NGC 6475.  
Membership is confirmed and a detailed abundance analysis is 
presented for four probable Ap members of this cluster, as well as for one normal B member.

\section{Membership Determination}

\subsection{Properties of NGC 6475}

NGC 6475 (M7) is a well-known open cluster in the constellation Scorpius.  
The cluster has a heliocentric distance of $280 \pm 26$ pc (Robichon  et al. 1999). A logarithmic age of $8.35 \pm 0.1$~y 
was determined by Meynet et al. (1993) using the cluster UBV colour-magnitude diagram. This cluster age was confirmed, 
and the age uncertainty derived, by fitting isochrones to the H-R diagram positions of the cluster stars.  Photometric data for 
the cluster stars was obtained from the WEBDA on-line database\footnote{{\tt http://www.univie.ac.at/webda/}} (Mermilliod, 1995).  
In general this list was not checked to confirm membership, however a few of the more important objects (eg. red giants) were investigated. 
Isochrones were generated by linear interpolation of the tabulated evolutionary model calculations of 
Schaller et al. (1992), for standard mass loss rate and a metallicity Z=0.020.   
With a focus on the 2 confirmed red giant members of NGC 6475, the age of the cluster 
was constrained to be $8.35 \pm 0.1$ logarithmic years.  As a comparison, Kharchenko et al. (2005) derive ages for nearly 200 clusters, 
including NGC 6475.  Their method involves fitting isochrones to individual cluster members and then determining an average 
cluster age.  Kharchenko et al. determined a logarithmic age of 8.22 for NGC 6475 and they quote a literature value 
(from a private communication by A. V. Loktin in 2004) of 8.48.  
These two values are in agreement with our values, within the total estimated error bar. 
 
There have been several modern studies of the cluster's apparent motion: van Leeuwen (1999) and Robichon et al. (1999) both used 
Hipparcos data to derive the cluster proper motion, while Dias et al. (2001) employed Tycho-2 data.  van Leeuwen (1999) reports 
proper motion in right ascension $\mu_\alpha = 2.64 \pm 0.35$ \may, proper motion in declination $\mu_\delta = -4.87 \pm 0.22$ \may, 
and parallax $\pi=3.71 \pm 0.32$ \mas.  Robichon et al. (1999) report 
$\mu_\alpha = 2.59 \pm 0.34$ \may, $\mu_\delta = -4.98 \pm 0.21$ \may, and $\pi =  3.57 \pm 0.30$ mas.  
Dias et al. (2001) report $\mu_\alpha = 1.7 \pm 2.1$ \may\, and $\mu_\delta = -3.7 \pm 2.1$ \may. These values are 
all in mutual agreement. Cluster radial velocity was investigated by Gieseking (1985), who found $ -15.3\pm 0.5$ \kms, 
and by Robichon et al. (1999), who found $-14.7 \pm 0.2$ \kms.

\subsection{Membership Data for the Stars}

 Five stars in NGC 6475 were analysed in this study: HD~162305 (B9), HD~162576 (B9p), HD~162725 (B9p), HD~320764 (A6), 
and HD 162817 (B9). The spectral classifications are from Renson's catalogue (1991). 
Proper motion and parallax data were found for HD 162576, HD 162725, and HD 162817 in the
Hipparcos catalogue (ESA 1997).  Additional proper motion data 
were found for all five stars in the Tycho-2 data base (H\o g et al. 2000).  Additional 
proper motion data for all five stars were found in Bastian and R\"oser's Positions 
and Proper Motions -  South catalogue (R\"oser \& Bastian, 1993). Although consistent with the results of the former studies, 
the PPM-S data are significantly less precise and will not be considered further.  The kinematic data for individual stars, 
which are consistent with the cluster motion discussed in Sect. 2.1, are 
reported in Table \ref{other_kin}.  

Radial velocities were reported by Gieseking (1985) for all 5 stars.  Additionally, radial 
velocities were measured from our own observed spectra.  During the synthetic spectrum fitting procedure radial 
velocity was considered to be a free parameter.  The quoted errors take into account noise in the observed 
spectrum, uncertainties in \vs, and unmodelled blended lines.  The measurements of Gieseking are in 
reasonable agreement with our own radial velocities, and all are consistent with the cluster radial 
velocities from Gieseking (1985) and Robichon et al. (1999)
  
Based on this analysis, we conclude that all five stars are high probability members of NGC 6475.

\begin{table*}
\begin{tabular}{cccccccc}
\hline \hline \noalign{\smallskip}
     & \multicolumn{2}{c}{Tycho-2}              & \multicolumn{3}{c}{Hipparcos}                                  & Gieseking       & This Work \\
\noalign{\smallskip} \hline \noalign{\smallskip}
Star &  $\mu_\alpha$ \may  &  $\mu_\delta$ \may  & $\mu_\alpha$ \may    &  $\mu_\delta$ \may  & $\pi$ \may       & $V_r$ \kms      &  $V_r$ \kms  \\
\noalign{\smallskip} \hline \noalign{\smallskip}
HD 162305 & $2.2 \pm 1.5$ & $-4.2 \pm 1.5$      & $ -- $               & $ -- $               & $--$             & $-12.3 \pm 1.6$ & $-19.0 \pm 2.6$ \\
HD 162576 & $3.1 \pm 1.2$ & $-3.6 \pm 1.3$      & $ 3.10\pm 1.33$      & $-4.92\pm 1.00$      & $3.94\pm 0.97$   & $-14.8 \pm 1.6$ & $-13.7 \pm 1.7$ \\
HD 162725 & $4.2 \pm 1.3$ & $-5.6 \pm 1.3$      & $ 4.33 \pm 1.09$     & $-4.59\pm 0.78 $     & $3.35\pm 0.90$   & $-9.4 \pm 1.2$  & $-14.4 \pm 2.6$ \\
HD 320764 & $0.7 \pm 1.5$ & $-5.6 \pm 1.5$      & $ -- $               & $ -- $               & $ --$            & $-12.0 \pm 2.4$ & $-19.1 \pm 3.3$ \\ 
HD 162817 & $3.2 \pm 1.3$ & $-4.5 \pm 1.3$      & $ 10.5\pm 4.1$       & $-4.62\pm 0.70$      & $3.87\pm 0.84$   & $-15.7 \pm 1.5$ & $-13.2 \pm 2.7$ \\
\noalign{\smallskip} \hline \noalign{\smallskip}
NGC 6475  & $1.7 \pm 2.1$ & $-3.7 \pm 2.1$      & $2.59 \pm 0.34$      & $-4.98 \pm 0.21$     & $3.57 \pm 0.30$  & $ -15.3\pm 0.5$ & $ -- $     \\
\noalign{\smallskip} \hline \hline

\end{tabular}
\caption[]{Kinematic data for all five stars from the Tycho-2, Hipparcos, and Gieseking's (1985) catalogues.  
 Tycho-2 kinematic data for the cluster is obtained from Dias et al. (2001), Hipparcos kinematic values are from Robichon et al. (1999). }
\label{other_kin}
\end{table*}

\section{Basic physical properties}

Three reddening-corrected methods were used to calculate the effective temperature of 
the four probable Ap stars.  uvby$\beta$ and Geneva photometry were obtained for each star from the 
General Catalogue of Photometric Data\footnote{{\tt http://obswww.unige.ch/gcpd/}}  
(Mermilliod at al. 1997).  A first determination of the effective temperature and surface gravity were obtained using TempLogG (St\"utz et al. 2002).  
This software package attempts to automatically correct for reddening.  
A second determination was obtained using Geneva photometry and the calibration of 
North and Nicolet (1990), assuming their [M/H](Z) = 1 coefficients.  Finally, the Hauck and North (1992) recalibration of 
the North and Nicolet (1990) calibration, adapted specifically for Ap stars, was used.  This gave a third effective 
temperature, although no corresponding \lgg\, value.  These methods make use of the reddening free {\it X} and {\it Y} parameters.  The 
resulting values from the three calibrations were then averaged to give 
the adopted temperature and surface gravity.  The standard deviations were adopted 
as the uncertainties of the average values.  The uncertainty values for our effective temperatures may be overly optimistic.  
We have implicitly assumed that the calibrations used are independent, and that there are no systematic errors in the resulting temperatures.  
Additionally, we have only sampled a small number of calibrations.  A more conservative estimate might be $\pm 500$ K in \teff.  
However, lacking evidence that our uncertainties are incorrect, we will proceed with the derived \teff\, values reported in Table \ref{basic_properties}.

For the normal B9.5 star HD 162817,  
TempLogG and the calibration of North and Nicolet (1990) were used to determine atmospheric properties.   
(The calibration of Hauck and North was not appropriate for this presumably chemically normal star.)   
As an additional check of the adopted \teff\, and \lgg\, 
values, synthetic ATLAS9 Balmer line profiles with corresponding temperatures and gravities were compared with the observed H$\alpha$ profiles. 
Good agreement was found. Spectrum synthesis for HD 162817 with these values also provides a better fit than 
models adjusted in \teff\, by 500 K or in \lgg\, by 0.5. In particular, lines produced by different ionization 
states of the same element could be more accurately fit at this temperature and \lgg. 

\begin{figure}
   \centering
   \includegraphics[width=3.4in]{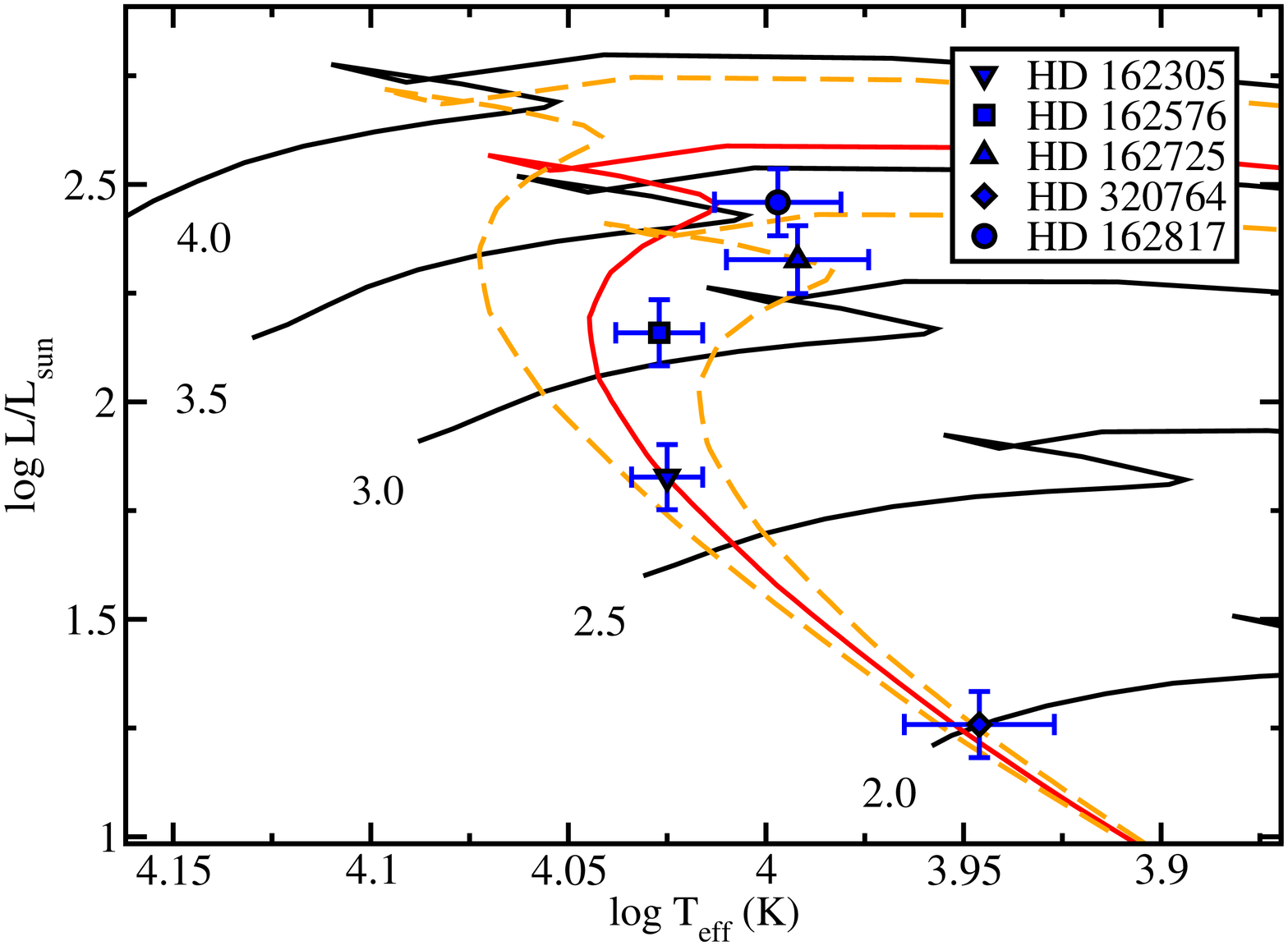}
   \caption{H-R diagram showing the derived positions of stars discussed in the paper, with evolutionary tracks and isochrones 
for log(age) = $8.35 \pm 0.1$, based on the calculations of Schaller et al. (1992) for standard mass loss and Z=0.02.  Evolutionary 
tracks are labelled by mass in $M_\odot$. }
   \label{h-r}
   \end{figure}

The stars were placed on an H-R diagram based on their effective temperature and bolometric 
corrected luminosity.  The cluster parallax from Robichon et al. (1999) and the bolometric correction of Balona (1994) 
were used to compute the luminosities.  The H-R diagram is presented in Figure \ref{h-r}, 
showing the derived positions of the stars as well as 
the cluster isochrone (8.35 log years) and evolutionary tracks from 2 to 4~$M_\odot$ (Schaller et al. 1992).  From the diagram 
we estimate masses for the five stars, which are shown in Table \ref{basic_properties}.  Radii for the stars 
are estimated from effective temperature and luminosity, and included in Table \ref{basic_properties}.  
The fraction of the main sequence lifetime elapsed for each star, $\tau$, is also given in Table \ref{basic_properties}.  
The $\tau$ values are calculated using the cluster's age and the evolutionary track corresponding to the star's mass, 
with uncertainties taking into account the uncertainty on the stellar mass and cluster age. Finally, we determined an age for 
each star based on its H-R diagram position.  These ages are fully consistent with the cluster age derived above. 
The uncertainties of these values are based on the range of isochrones that 
pass through the rectangle described by the star's position error bars.  Departures from standard mass loss and Z=0.02 are not considered.  
While the error bars on some of the stars are small, one must note that these objects are nearing the 
terminal age main sequence (TAMS) line, where the isochrones strongly diverge.  
Thus, for some of our stars, we derive an unusually precise age based on H-R diagram position. Despite this relative precision, 
we consider the cluster age more definitive, and use it where possible.
If we were to use a more conservative uncertainty on \teff\, of 500 K, the uncertainties on 
$\log\, L/L_\odot$ would increase by about 0.01, mass would be approximately 0.05 $M_\odot$ more uncertain, radius 0.1 $R_\odot$ more uncertain, 
and $\tau$ 0.05 more uncertain.  The H-R diagram age would be 0.05 log yr more uncertain for our more 
precise limits, up to 0.2 log yr more uncertain for our least precise values.

In principle, the cluster age can be used to constrain the H-R diagram location of cluster members, providing accurate masses and luminosities.  
However, the divergence of isochrones toward the TAMS leads to large uncertainties 
in the derived properties of more massive cluster members using this method.  In the case of our 
sample of five stars this method did not provide any advantage.  The H-R diagram positions of the 
stars are already well constrained, thanks to the relatively precise cluster parallax.  
While no more precise, the luminosities and masses determined using the cluster age are in agreement with the 
values presented in Table \ref{basic_properties}. 

\begin{table*}
\begin{tabular}{cccccccc}
\hline \hline \noalign{\smallskip}
Star & Temperature (K)       & \lgg        & $\log\, L/L_\odot$ & Mass ($M_\odot$)& Radius ($R_\odot$) & $\tau$    & H-R Age (log yr)\\	  
\noalign{\smallskip} \hline \noalign{\smallskip}								      				  
HD 162305 & $10580 \pm 220$ & $3.9 \pm 0.1$ & $1.827 \pm 0.075 $ & $2.7 \pm 0.1$ & $ 2.5 \pm 0.2$ & $0.5 \pm 0.1$  & $\leq 8.45$ \\		  
\noalign{\smallskip}												      				  
HD 162576 & $10640 \pm 270$ & $3.7 \pm 0.1$ & $2.159 \pm 0.076 $ & $3.1 \pm 0.2$ & $ 3.6 \pm 0.3$ & $0.7 \pm 0.2$  & $8.40^{+0.05}_{-0.04}$ \\
\noalign{\smallskip}												      				  
HD 162725 & $9820 \pm 400$  & $3.5 \pm 0.2$ & $2.327 \pm 0.078 $ & $3.2 \pm 0.3$ & $ 5.1 \pm 0.5$ & $0.7 \pm 0.2$  & $8.43^{+0.11}_{-0.06}$ \\
\noalign{\smallskip}												      				  
HD 320764 & $8820 \pm 390$  & $4.2 \pm 0.1$ & $1.258 \pm 0.076 $ & $2.0 \pm 0.1$ & $ 1.9 \pm 0.2$ & $0.2 \pm 0.1$  & $8.35^{+0.10}_{-0.25}$ \\		  
\noalign{\smallskip}												      				  
HD 162817 & $9940 \pm 360$  & $3.4 \pm 0.1$ & $2.459 \pm 0.077 $ & $3.4 \pm 0.2$ & $ 5.8 \pm 0.6$ & $0.8 \pm 0.2$  & $8.43^{+0.05}_{-0.08}$ \\
\noalign{\smallskip}
\hline \hline

\end{tabular}

\caption[]{Derived physical properties of the five program stars.  $\tau$ is the time fraction of the star's 
main sequence evolution that has been completed, assuming the cluster age for each star and evolutionary tracks 
from Schaller et al. (1992) with Z = 0.02.  The H-R age is the age estimated for each star individually, based only on its 
H-R diagram position. }
\label{basic_properties}

\end{table*}

\section{High-Resolution Spectra}

The observed spectra employed in this investigation were obtained within the context of the 
Paranal Observatory Project (POP) program (ESO DDT Program ID 266.D-5655), using the UVES high-resolution spectrograph at ESO-VLT.  
This project aims to collect a library of high resolution spectra which are readily available to the 
public (Bagnulo et al., 2003).  A large number of field stars 
have been observed by the UVES-POP, as well as stars in two open clusters: NGC 6475 and IC 2391.  
In total about 80 cluster stars and about 330 field stars have been observed.

The spectra cover about 300 nm to 1000 nm, with two small gaps, at a resolving power of approximately 80000. 
The majority of the data reduction was performed by an automated pipeline developed for the UVES instrument.  
Parameters were tailored for the POP, using the ``average extraction'' method instead of the ``optimal extraction'' method. 
This pipeline automatically considers bias and flat field corrections, and wavelength calibration.  
The pipeline also attempts to correct for bad CCD pixels and subtracts the 
background sky from the spectrum.  For a further description of the pipeline see Ballester et al. (2000).  
Once downloaded from the POP internet site, the individual spectral orders were 
continuum normalized with the aid of the {\tt continuum} routine in the Image Reduction and Analysis Facility (IRAF)\footnote{IRAF 
is distributed by the National Optical Astronomy Observatory, which is operated by the Association of Universities for Research in 
Astronomy (AURA), Inc., under cooperative agreement with the National Science Foundation.} using a 5th to 10th order Legendre polynomial 
depending on the order of the spectrum.  
The individual orders were then merged, and the correction to the heliocentric rest frame was calculated using the IRAF {\tt rvcorrect} routine.  
The UVES POP spectra are provided in both merged and unmerged `2D' forms, however we found the unmerged form to be more useful.  
The merged spectra display an artifact, apparently due to the merging process: there is a slight ripple in the continuum level.  
Thus we prefer the unmerged `2D' spectra which contain the original unmerged echelle orders.

\section{Spectrum Synthesis}

The observed spectra were modelled using the Zeeman2 (Landstreet 1988; Wade et al. 2001) spectrum synthesis code.  
Chemical abundances and projected rotational velocities \vs\, were determined by direct fitting of the 
synthesized spectra.  For each of the five stars, several spectral regions were 
examined. Uniform vertical and horizontal abundance distributions were assumed.  While 
a uniform horizontal distribution is almost certainly not the case for some elements in some stars, 
with only one observation per star, it is impossible to accurately model surface distributions.  

For all Ap/Bp stars, a dipolar photospheric magnetic field was assumed, with an intensity of 1 kG, oriented with the positive magnetic pole 
pointing directly along the line-of-sight toward the observer.   No positive detections of the magnetic 
fields of the Ap star members of NGC 6475 are available in the literature.  Bagnulo et al. (2006) 
obtain null results for the longitudinal fields of HD 162305, HD 162725 and HD 320764 with 1$\sigma$ error bars of about 50 G. 
However, the longitudinal magnetic fields of Ap stars are variable, and have a typical rms of only about 300 G (Bohlender \& Landstreet 1990). 
Given the small number of magnetic observations 
(just one for HD 162725 and HD 320764, and two for HD 162305) the stars could quite probably have been observed when the 
visible field configuration produced a longitudinal magnetic field below the detection threshold.  
We therefore assume that these are ``typical" magnetic Ap stars with surface field strengths of about 1 kG.
Inclusion of this weak magnetic field introduces magnetic desaturation, and therefore decreases 
slightly (by about 0.05 dex) the abundances inferred from most lines.

For the normal star HD 162817 it was assumed that there was no magnetic field.  
Microturbulence was also assumed to be absent in the atmosphere of this star. 
While this may not be strictly correct for a star with \teff\, 10000 K, there are no discrepancies between our synthetic and observed spectra 
of the type one would expect if there were a significant microturbulence.  Both weaker and stronger lines of the same species 
are well fit with a single abundance value.  However, due to the higher projected rotational 
velocities of this star and our program stars, few (if any) weak lines (on the linear part of the curve of growth) 
are actually detected and modelled.  Therefore undetected microturbulence may exist, but would have an impact, at most, 
at the level of our uncertainty in abundance.  For example a 1 km/s microturbulence would decrease our abundance values 
for this star by at most 0.1 dex.  

For all program stars except HD~320764, microturbulence was assumed to be suppressed, either directly or indirectly, 
by the presence of a magnetic field.  Therefore, no microturbulence was assumed in computing the line 
profiles of HD~162305, HD~162576, and HD~162725.  For HD~320764 a 1 \kms\, microturbulence was assumed. 
In this star, it proved impossible to determine a microturbulence observationally.  The very large rotational 
broadening makes finding the necessary mix of weak and strong lines of the same species impossible. 
However, for a normal A star with a temperature of 8800 K we expect there to be a small microturbulence (Landstreet 1998), 
so a conservative value of 1 \kms\, was chosen. 

Input atomic line data was extracted from the Vienna Atomic Line Database (Kupka et at. 1999).  An ``extract stellar'' 
request was used with the \teff\, and \lgg\, values from Table \ref{basic_properties}.  
1 dex overabundances of Si, Cr, Fe, and 3 dex overabundances all rare earth elements were used to ensure a complete line list.  
A 2 \kms\, microturbulence was used to simulate a weak magnetic field.  Only lines with a central line depth greater then 0.01, 
as a fraction of the continuum, were used.  The list of lines and atomic data employed is provided in Table 3 (available only in electronic form). 

The general spectrum fitting strategy involved iteratively synthesizing a spectrum, examining the fit by eye, 
and then adjusting the input model parameters accordingly.  In some more complicated cases, for stars with line profile 
shapes which depart significantly from the rotationally-broadened model (probably due to surface abundance non-uniformities), 
correspondence between model and observed equivalent widths was also verified.  
Uncertainties on the final \vs\, and abundance values are based on the dispersion of values determined from 
many absorption lines.  If few absorption lines of an element are present in the observed spectrum, the uncertainty includes 
a conservative estimate for the change in abundance necessary to shift the synthesized spectrum well above the noise level, 
and any possible normalization errors.  
Thus, uncertainty values presented from spectrum synthesis correspond to a $2-3\sigma$ confidence level.

\subsection{HD 162817}

\begin{figure*}
   \centering
   \includegraphics[]{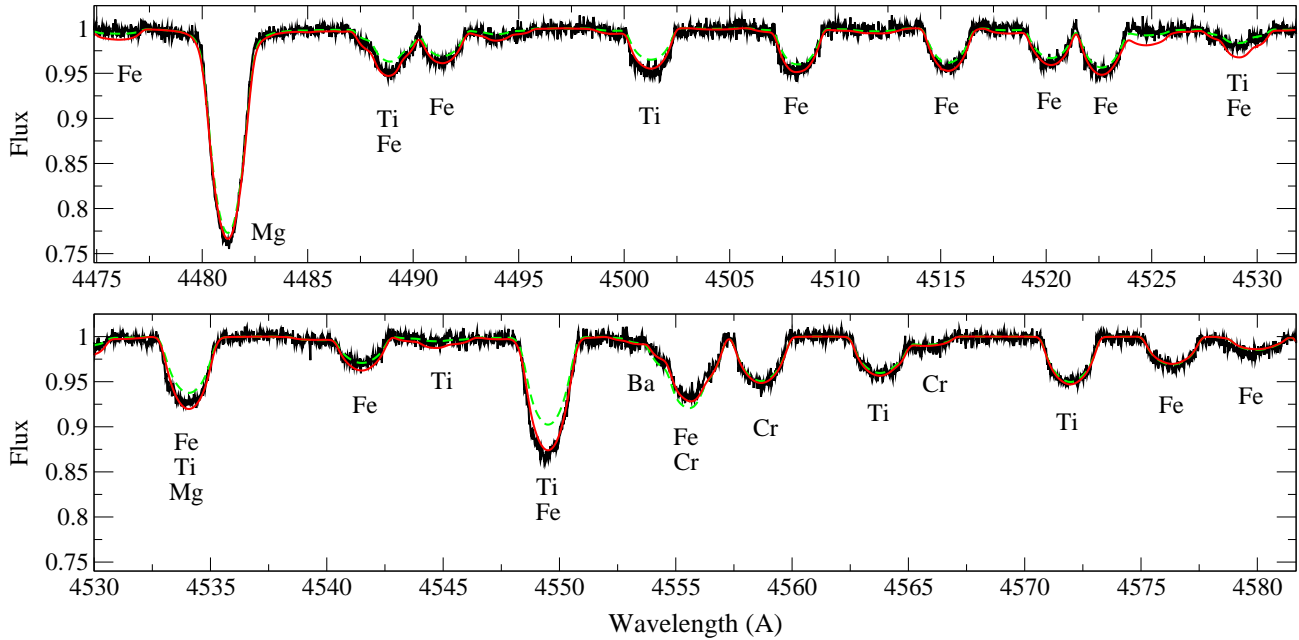}
   \caption{A sample segment of spectrum from HD 162817.  The black line is the observed spectrum, 
     the solid smooth line is the calculated best fit spectrum.  The dashed smooth line is a calculated 
     spectrum assuming solar abundances.  The major contributors to each absorption feature have are indicated. }
   \label{162817_spec}
   \end{figure*}

The primary motivation for examining HD 162817 was to check that our fitting procedure produced the expected, 
approximately solar, abundance values for a normal late B-type star.  
HD 162817 is a normal B9 star with no known peculiarities and V magnitude of 6.11.  The observed 
spectrum for this star is of good quality, with a peak signal-to-noise ratio of approximately 280.  
The line profiles of this star are moderately broad, and profiles of unblended lines show no obvious asymmetries.  
A solar abundance ATLAS9 model atmosphere (Kurucz 1993) with a temperature of 10000 K and \lgg\, of 3.5 was 
used to model the spectrum of this star.  A projected rotational velocity (\vs) of $79 \pm 3$ \kms was found.  
The mean best fit chemical abundances, derived for 12 elements, are reported in Table \ref{final_ab}.  
Illustrative regions of the observed spectrum, compared with the best fit synthetic spectrum, are shown in Figure \ref{162817_spec}.  
For most elements, the derived abundances of HD~162817 are consistent with solar values.  In particular, iron peak elements such as Cr, 
normally overabundant in Ap stars, display solar abundances.  This is not a surprising result, but it confirms the accuracy of our fitting method.  

Hempel \& Holweger (2003) determined abundances for a number of late B stars including HD 162817.  They first determined the 
effective temperature \teff=9190 K and \lgg=3.15, based on Str\"omgren photometry and the calibration of Napiwotzki et al. (1993) 
and assumed a microturbulent velocity of $1 \pm 1$ \kms.  
Notably, this effective temperature is not consistent with the spectral classification of B9.  They derived LTE 
abundances, in dex relative to the sun, of O (1.1), Mg (0.06), Si (0.12), Ca (-0.53), Fe (0.29), and Sr (-0.07).  Our abundances 
for Mg, Si, and Fe are consistent with theirs within $2\sigma$.  We derive no Sr abundance for comparison.  
Our Ca abundance ($0.24 \pm 0.4$ dex above solar) is greater than that of Hempel \& Holweger by $\sim 0.7$ dex.  
Hempel \& Holweger provide no uncertainty on their value, but if their uncertainty is similar to ours, 
then the values may in fact be consistent within $2\sigma$.  Additionally, their 
Ca abundance value is derived from only one line, reducing our confidence in the value somewhat.  
Hempel \& Holweger use Ca atomic data extracted from VALD, as do we, so the log {\it gf} values used should be the same.
The Ca II K line at 3934~\AA\, that Hempel \& Holweger use has been known to display the effects of stratification 
in many Ap stars (Babel, 1994),  however as HD~162817 appears to be a normal B star this not likely to be the source of the discrepancy.  
Additionally, the Ca II K line is a particularly strong line, which leaves the slight possibility of some 
undiagnosed non-LTE effect reducing the accuracy of an abundance based solely on this line.  

Hempel \& Holweger, using the O~{\sc i}~$\lambda 7771$ triplet, derive a non-LTE abundance of 0.63 dex above solar.  
This is well above our abundance of solar $\pm 0.1$ dex, based on O~{\sc i}~$\lambda 6162$.  To investigate this discrepancy, we redetermined our 
abundances using Hempel \& Holweger's temperature and \lgg\, resulting in an increase in our abundance by about 0.2 dex.  We then determined 
an abundance for the O~{\sc i}~$\lambda 7771$ triplet (assuming LTE) and obtained an abundance 1.3 dex over solar.  
There appears to be a large NLTE effect in this feature, likely responsible for much of the discrepancy we observe.  
If this were fully taken into account, along with the correct temperature, then our results would likely be much more similar.

\subsection{HD 162576}

\begin{figure*}
   \centering
   \includegraphics[]{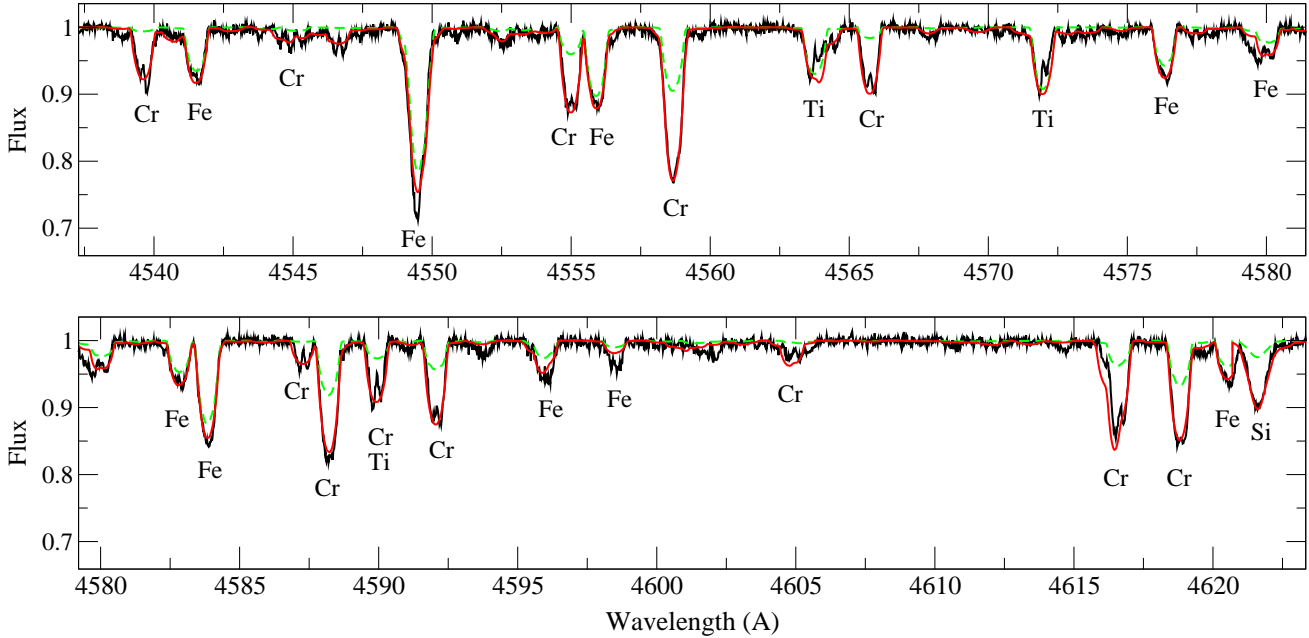}
   \caption{A sample segment of spectrum from HD 162576.  The lines and labels have the same significance as in Figure \ref{162817_spec}.  }
   \label{162576_spec}
   \end{figure*}

HD 162576 is classified as B9p SiCr, with a V magnitude of 6.99 and has an approximate rotation period of 3.43 days (Catalano \& Renson 1998).  
The catalogue of Renson et al. (1991) indicates that this is a highly probable chemically peculiar star.  The observed spectrum 
for HD 162576 is of good quality, with a peak signal-to-noise ratio of 300, and sharp lines that are relatively smooth and unblended.  
Unfortunately a segment of the spectrum in the red, from about 
5750~\AA\, to about 6700~\AA, was of unusually poor quality and hence unavailable for the purposes of modelling.  
A temperature of 10500 K and a surface gravity of 4.0 were adopted for the solar abundance ATLAS9 model atmosphere of HD 162576.  
We derived a projected rotational velocity of $28\pm 3$~\kms\, and mean chemical abundances for 17 elements, reported in Table \ref{final_ab}.
A sample segment of spectrum with its best fit model is shown in Figure \ref{162576_spec}.  
Clear overabundances of Si, Cr, and Mn can be seen while O, Mg, and Ca are underabundant.  The cores of both the 
Cr and Ti lines in this spectrum show some structure.  Ti lines are significantly deeper on the blue side of the line then the red.
The Cr lines display a slight rise at the very centre of the line.  This subtle structure is likely the result of non-uniform 
surface abundance distributions of these elements, and their modelling is beyond the scope of this paper.

\subsection{HD 162725}

\begin{figure*}
   \centering
   \includegraphics[]{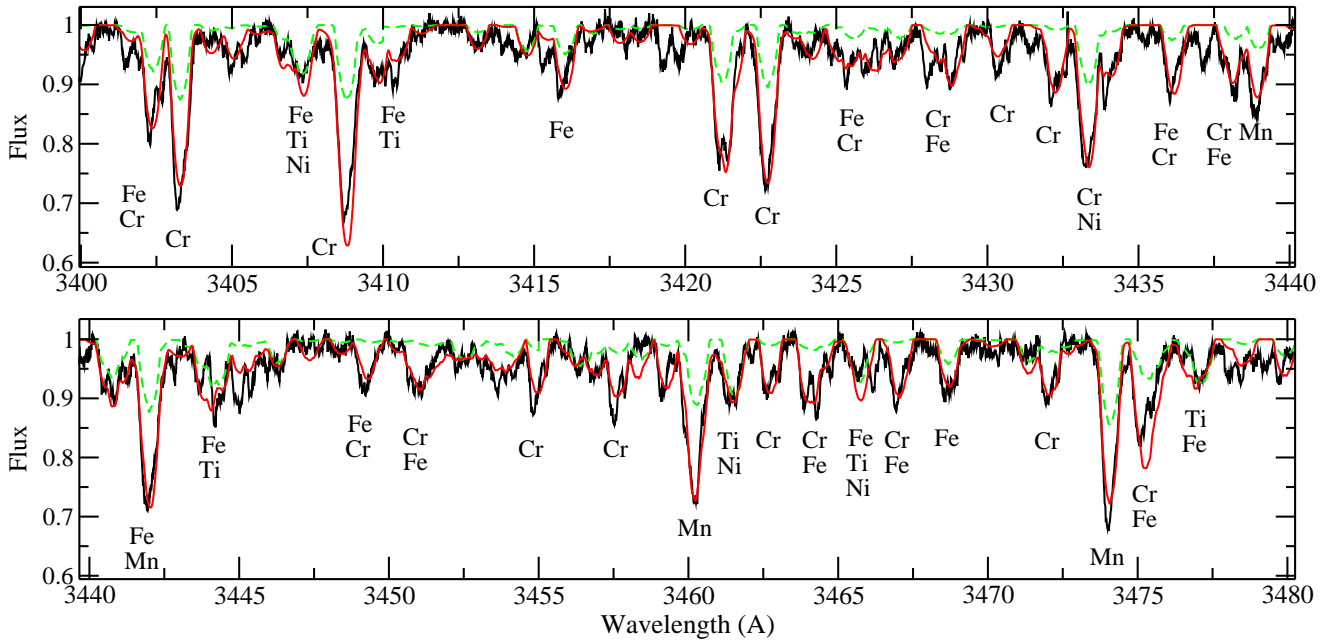}
   \caption{A sample segment of spectrum from HD 162725.  The lines and labels have the same significance as in Figure \ref{162817_spec}. }
   \label{162725_spec}
   \end{figure*}

HD 162725 is a B9p SiCr star with a V magnitude of 6.42 and a rotation period of 4.459 days (Renson \& Catalano 2001).  
Renson et al. (1991) list this star as a well-established chemically peculiar star.  This is another sharp-lined 
star with a high (320) signal-to-noise ratio. Line profiles of some elements show  
obvious line profile structure and asymmetry.  A temperature 
of 9750 K and a surface gravity of 3.5 were adopted for the model atmosphere.  
One notable feature of the spectrum is a distinct asymmetry in the chromium lines.  There is a 
clear depression on the blue side of the line and a peak near line centre.  This suggests 
that there may be a strongly inhomogeneous distribution of Cr across the surface of this star. 
Our uniform-abundance model provides a relatively poor fit to these complex profiles. In order to 
derive the abundance of Cr, we verified that the equivalent widths of the observed and synthesized spectral lines were in good agreement.  
This technique of matching equivalent widths was used for Fe as well, as lines of this element also appear
to display slightly asymmetric line profiles.  The resulting mean abundances for 15 elements are reported in Table \ref{final_ab}.
A sample section of spectrum is shown in Figure \ref{162725_spec}.  
HD 162725 exhibits strong overabundances of iron peak elements, as well as some rare earth elements such as Nd and Eu.  
Si appears overabundant, however O, Mg and Ca are under abundant.

\subsection{HD 162305}

HD 162305 is classified as a B9 star with a V magnitude of 7.81.  Renson et al. (1991) 
list this star as having a doubtful chemically peculiar nature.  It was included in our study, 
as a likely Ap star, due to its large $\Delta a$ photometric value (Maitzen \& Floquet, 1981) and unusual line profiles.  
Our observation of this star has a very good peak signal-to-noise ratio of 420.
HD 162305 has broader lines then the previous 2 stars, and again asymmetries are clearly 
present in the high-resolution spectrum.  The strong asymmetries are most obvious in the Cr lines, there is a large flux increase 
in the centre of the line profile and both blue and red wings are significantly deeper.  Other elements 
also display asymmetries: Fe lines tend to be deeper in the blue wing and Ti lines rise significantly in the centre of the core.   
An effective temperature of 10500 K and surface gravity of 4.0 were adopted for the ATLAS9 model atmosphere. The notably broader lines of
this star imply a significantly greater projected rotational velocity.  
The calculated \vs\, of $85 \pm 5$ \kms is roughly three times larger for this star than for the previous two.  
Since the total amount of light absorbed is independent of rotational broadening, the line depth will decrease 
as the line width grows.  As a result, line cores in the spectrum of HD 162305 are closer to the noise level.  Additionally, any asymmetries in 
the line profile, due to inhomogeneous distributions of elements across the surface of the star, will be much more prominent in 
the spectrum of a faster rotating star.  These two effects combine to increase the difficulty of the fitting procedure for HD 162305.  

\begin{figure*}
   \centering
   \includegraphics[]{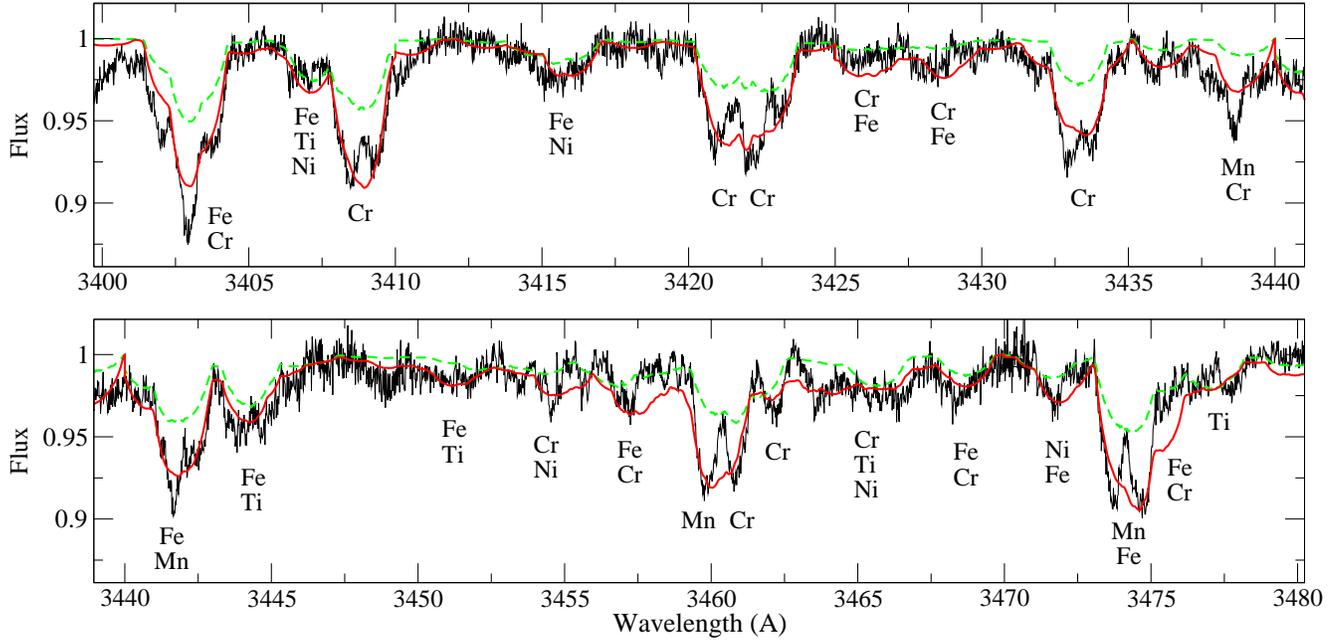}
   \caption{A sample segment of spectrum of HD 162305.  The lines and labels have the same significance as in Figure \ref{162817_spec}. }
   \label{162305_spec}
   \end{figure*}

Fitting \vs\, was accomplished by focusing on the wings of a large number of lines; the core was 
largely neglected due to line profile structure.  In order to 
determine element abundances, we once again fit the equivalent widths of lines with weak blending.  
The best fit abundances are reported in Table \ref{final_ab} and a sample 
section of spectrum is shown in Figure \ref{162305_spec}.  
In the photosphere of HD 162305, Cr and Mn are strongly overabundant, as is Nd, while O and Mg are underabundant.

\subsection{HD 320764}

\begin{figure*}
   \centering
   \includegraphics[]{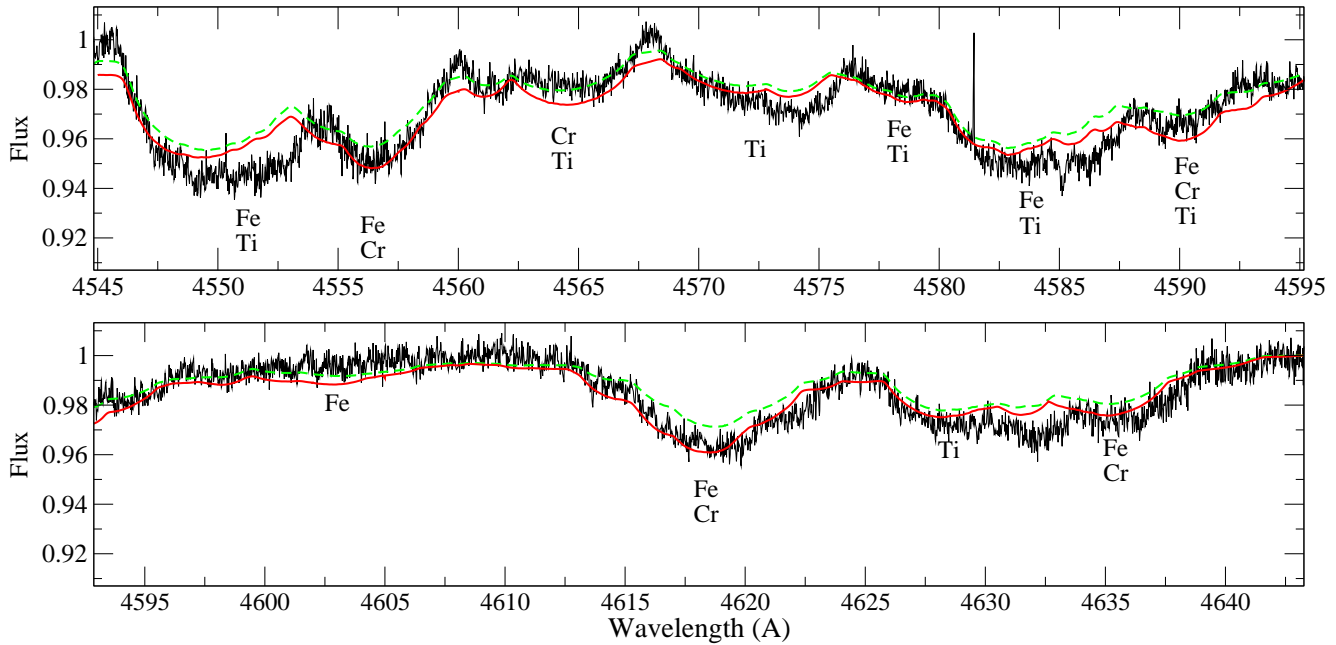}
   \caption{A sample segment of spectrum from HD 320764.  The lines and labels have the same significance as in Figure \ref{162817_spec}.  }
   \label{320764_spec}
   \end{figure*}

The fifth star, HD~320764, is classified as A6, and has a V magnitude of 8.93.  
Renson et al. (1991) consider this star to be of a doubtful chemically peculiar nature.  This star 
was included in our study, as possibly peculiar, do to its larger $\Delta a$ photometric value (Maitzen \& Floquet, 1981).
The spectrum of HD~320764 exhibits the broadest lines and a signal-to-noise ratio of 320.  Despite the high signal-to-noise ratio, 
the shallow lines of this star make noise more of a problem then in the other stars investigated.  
A temperature of 8750~K and surface gravity of 4.0 were adopted for the model atmosphere.   A 1 \kms\, microturbulence was used in the 
spectrum synthesis.  If we were to increase this value to 2 \kms, this would result in a decrease in most abundances by about 0.2 dex.   
The \vs\, determined by spectrum synthesis is 225 \kms . Such a large \vs\, makes modelling very 
challenging due to the very shallow and heavily blended lines.  In the spectrum of a star with lines this broad 
and shallow, continuum normalization becomes a large potential source of error
and must be dealt with carefully.  No clear asymmetries are seen in this spectrum.  The fitting 
procedure focused on both the core and wings of the lines, although in this case drawing a 
distinction between wing and core is not very useful.  It is very unusual for an Ap star to have a \vs\, 
value this large.  
Modelling of the spectrum of this star shows near solar abundances for Cr, Fe, Ti and Ni with no clear 
overabundances of rare earth elements.   The abundance of Ca appears marginally greater than solar, 
however only three lines were used, all of which are particularly sensitive to microturbulence.  
If we were to use a microturbulence of 2 \kms\, the abundance would fall to $5.3 \pm 0.5$ dex, 
well within uncertainty of solar.  Due to this sensitivity and the uncertainty in the microturbulence of HD~320764, 
the Ca abundance presented here is somewhat more uncertain then that of other elements.  
Although uncertain, the abundances derived for HD~320764 provide no evidence to  
suggest that the star is chemically peculiar. The derived abundances 
are reported in Table \ref{final_ab} and a sample section of spectrum is shown in Figure \ref{320764_spec}.

   \begin{figure}
   \centering
   \includegraphics[width=6.0cm]{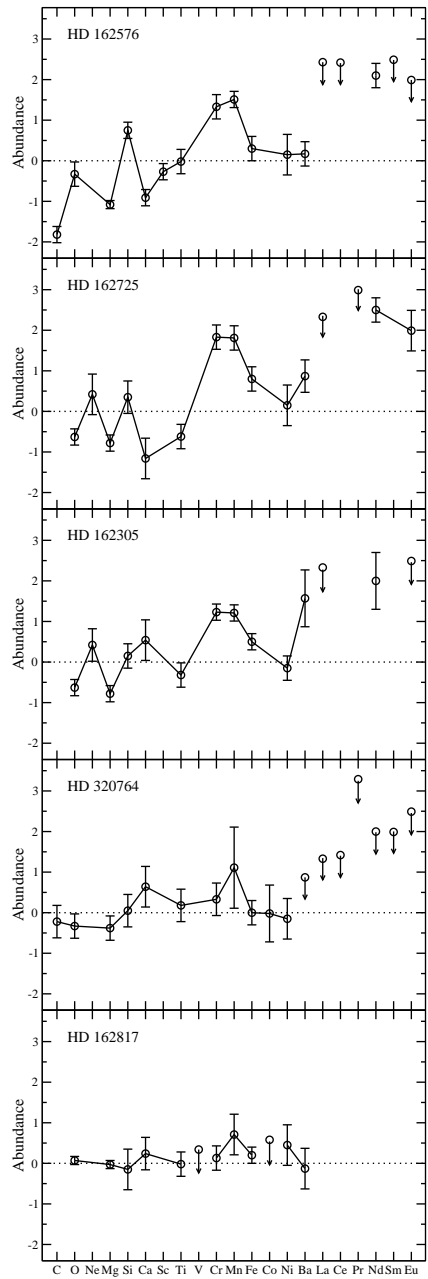}
   \caption{Abundances, relative to the sun, for all the modelled elements in the five stars.}
   \label{total_ab}
   \end{figure}

\setcounter{table}{3}
\begin{table*}
\centering
\begin{tabular}{ccccccc}
\hline \hline \noalign{\smallskip}
        & HD 162576       & HD 162725          & HD 162305          & HD 320764          & HD 162817          & Solar \\
\noalign{\smallskip} \hline \noalign{\smallskip}
 \teff\, (K) & $10640 \pm 270$ & $9820 \pm 400$ & $10580 \pm 220 $  & $ 8820 \pm 390 $   & $ 9940 \pm 360 $   & \\
 Mass ($M_\odot$) & $3.0 \pm 0.4$ & $3.3 \pm 0.5$ & $2.7 \pm 0.3$   & $2.0 \pm 0.3$      & $3.3 \pm 0.4$      & \\
\noalign{\smallskip} \hline \noalign{\smallskip}
 \vs\,  (\kms)   &  28  $\pm$ 3   &  32  $\pm$ 5   &  85  $\pm$ 5   &  225  $\pm$ 10     &  79  $\pm$ 3       & \\ 
\noalign{\smallskip} \hline \noalign{\smallskip}
 C   &  -5.3  $\pm$  0.2  &                    &                    &  -3.7  $\pm$  0.4  &                    & -3.48 \\ 
 O   &  -3.5  $\pm$  0.3  &  -3.8  $\pm$  0.2  &  -3.8  $\pm$  0.2  &  -3.5  $\pm$  0.3  &  -3.1  $\pm$  0.1  & -3.17\\ 
 Ne   &                   &  -3.5  $\pm$  0.5  &  -3.5  $\pm$  0.4  &                    &                    & -3.92\\ 
 Mg   &  -5.5  $\pm$ 0.1  &  -5.2  $\pm$  0.2  &  -5.2  $\pm$  0.2  &  -4.8  $\pm$  0.3  &  -4.45  $\pm$  0.1 & -4.42\\ 
 Si   &  -3.7  $\pm$  0.2 &  -4.1  $\pm$  0.4  &  -4.3  $\pm$  0.3  &  -4.4  $\pm$  0.4  &  -4.6  $\pm$  0.5  & -4.45\\ 
 Ca   &  -6.55  $\pm$ 0.2 &  -6.8  $\pm$  0.4  &  -5.1  $\pm$  0.5  &  -5.0  $\pm$  0.5  &  -5.4  $\pm$  0.4  & -5.64\\ 
 Sc   &  -9.1  $\pm$  0.2 &                    &                    &                    &                    & -8.83\\ 
 Ti   &  -7.0 $\pm$  0.3  &  -7.6  $\pm$  0.3  &  -7.3  $\pm$  0.3  &  -6.8  $\pm$  0.4  &  -7.0  $\pm$  0.3  & -6.98\\ 
 V    &                   &                    &                    &                    & $\leq$ -7.7        & -8.04\\
 Cr   &  -5.0 $\pm$  0.3  &  -4.5  $\pm$  0.3  &  -5.1  $\pm$  0.2  &  -6.0  $\pm$  0.4  &  -6.2  $\pm$  0.3  & -6.33\\ 
 Mn   &  -5.1  $\pm$  0.2 &  -4.8  $\pm$  0.3  &  -5.4  $\pm$  0.2  &  -5.5  $\pm$  1.0  &  -5.9  $\pm$  0.5  & -6.61\\ 
 Fe   &  -4.2  $\pm$  0.3 &  -3.7  $\pm$  0.3  &  -4.0 $\pm$   0.2  &  -4.5  $\pm$  0.3  &  -4.3  $\pm$  0.2  & -4.50\\ 
 Co   &                   &                    &                    &  -7.1  $\pm$  0.7  & $\leq$ -6.5        & -7.08\\ 
 Ni   &  -5.6  $\pm$  0.5 &  -5.6  $\pm$  0.5  &  -5.9  $\pm$  0.3  &  -5.9  $\pm$  0.5  &  -5.3  $\pm$  0.5  & -5.75\\ 
 Ba   &  -9.7  $\pm$  0.3 &  -9.0  $\pm$  0.4  &  -8.3  $\pm$  0.7  & $\leq$  -9         &  -10.0  $\pm$  0.5 & -9.87\\ 
 La   & $\leq$ -8.4       &$\leq$  -8.5        & $\leq$ -8.5        & $\leq$  -9.5       &                    & -10.83\\ 
 Ce   & $\leq$ -8         &                    &                    & $\leq$  -9         &                    & -10.42\\ 
 Pr   &                   &$\leq$  -8.3        &                    & $\leq$  -8         &                    & -11.29\\ 
 Nd   &  -8.4  $\pm$  0.3 &  -8.0  $\pm$  0.3  &  -8.5  $\pm$  0.7  & $\leq$  -8.5       &                    & -10.50\\ 
 Sm   & $\leq$ -8.5       &                    &                    & $\leq$  -9         &                    & -10.99\\ 
 Eu   & $\leq$ -9.5       &  -9.5  $\pm$  0.5  & $\leq$  -9         & $\leq$  -9         &                    & -11.49\\ 
\noalign{\smallskip} \hline \hline
\end{tabular}

\caption[]{Derived rotational velocities and abundances, averaged over all windows, for the five stars studied.  
 Abundances reported for each element as $\log N/N_{\rm tot}$.  Quoted uncertainties in \vs\, and abundance can be 
 considered upper and lower limits, at about $2\sigma$.  Solar abundances are from Grevesse \& Sauval 1998.}
\label{final_ab}
\end{table*}

\section{Discussion and Conclusions}

Chemical abundances as well as projected rotation velocity have been determined for five 
confirmed members of NCG 6475 (M7).  The derived chemical abundances are presented graphically in Figure \ref{total_ab}.  
Three stars are clearly classical Ap/Bp stars, and the normal B9 star has been confirmed to have solar abundances.  
However, the supposed Ap star HD~320764 appears to be a chemically normal A star.  Abundances and abundance upper limits have been found 
for a wide range of elements from C to Eu, with typical $2\sigma$ uncertainties (generally derived from the dispersion of abundances 
determined from multiple absorption lines) of 0.3 dex. 

The chemically normal B star HD~162817 displays abundances that are within uncertainty of solar for essentially all elements.  
The three confirmed Ap stars all show strong overabundances (as compared to the sun and as compared to HD~162817)
of Cr and Mn.  Only marginal overabundances are found for Si, Fe and Ni, while Ti is 
solar or marginally underabundant.  The rare earths Nd and Eu are also overabundant when detected in the Ap stars.  
The light elements O and Mg are slightly underabundant.
HD~320764 displays abundances that are within uncertainty of solar for all elements detected, except for Ca and Mn.  
Mn is nearly within uncertainty of solar and based only on one spectral window.  Ca is also nearly 
within uncertainty of solar and the lines used are particularly sensitive to the choice of microturbulence, 
making the abundance somewhat less certain than that of other elements studied.
  
The structure and strong asymmetries observed in many lines of HD~162576, HD~162725 and HD~162305 support the 
conclusion that these stars are magnetic Ap/Bp stars.  Complex line profiles including many 
double-peaked profiles imply complicated surface abundance distributions.  
The abundances presented in this paper therefore represent averages of these complex distributions 
over the observable hemisphere of star at the time of observation.  Additional high 
resolution spectra of these three stars are necessary if the surface distributions are to be mapped.  

The overabundances of Cr and Eu are marginally greater for the cooler Ap star HD~162725 than for the two hotter Ap stars, 
consistent with the cooler temperatures seen in SrCrEu Ap stars when compared to Si Ap stars in the field (Wolff 1967).  
The values of Cr overabundance seen, and the trend with temperature, 
are consistent with the pattern observed by Ryabchikova et al. (2004). 
There is no clear relationship between overabundance of elements and rotation within the three chemically peculiar stars.  
However, it is notable that the chemically normal stars on average have significantly higher \vs\, then the Ap stars.  
This is consistent with published results, for example Abt \& Morrell (1995). 

HD~162576 and HD~162725 have identical age and initial composition, as well as the same \vs, mass, 
and their derived effective temperatures are only marginally different ($820 \pm 483$~K). 
A period analysis shows that the stars have similar periods at 3.43 and 4.459 days respectively 
(Catalano \& Renson 1998 and Renson \& Catalano 2001).  With our adopted uncertainties, the abundances of most 
elements are very similar in these two stars.  There are some marginally detectable differences: Si appears to be less abundant 
in HD~162725 then in HD~162576, while Ba is more abundant.  Cr and Fe may be more abundant in HD~162725 as well.  The similarities in derived 
abundances between these two stars suggest that age, mass, rotation speed, and possibly magnetic field strength may be enough to uniquely 
determine a set of chemical peculiarities present in a late-B star.  
The slight differences in abundance between the two stars may be related to the difference 
in rotation period, phase of rotation, or possibly magnetic field, between the two stars.  

HD~162305 and HD~162576 have identical age, initial composition, effective temperatures, and masses, but \vs\, differs by 
more than a factor of 3. These stars display many similarities, but there are some significant abundance differences 
(e.g. $1.45 \pm 0.5$ dex for Ca, $1.4 \pm 0.8$ dex for Ba).  Assuming that the true rotational velocities of these stars 
differ by a similar factor, it could be that these differences reflect the effects of rotational mixing. 

HD~162305 is also nearly identical to HD~162725 except in \vs.  Both stars have identical age and initial chemical composition, 
as well as very similar masses and temperatures.  There are many similarities between these two stars, as with HD~162576, 
but there are also some significant differences, particularly the $1.7 \pm 0.6$ dex difference in Ca, and smaller differences 
in Cr and Mn.  A study of the rotation periods of the five stars in this paper would be very useful, 
as it would allow for much more definite comments on the effect of rotational mixing in this sample of stars.  

Even more remarkably, HD~162305 and HD~162817 (our chemically normal comparison star) 
have formally identical \vs\, and only marginally different effective temperatures, 
as well as identical age and initial chemical composition. However, these stars display strong 
differences in the photospheric abundances of chemical elements (e.g. $1.1 \pm 0.4$ dex for Cr, $1.7 \pm 0.9$ dex for Ba). 
Presumably the difference results from the presence of a magnetic field, but this begs the question: what distinguishes 
the presence of magnetic field in co-eval, co-environmental stars with similar masses, temperatures and rotation speeds?

Some major questions about Ap stars remain.  In particular how their magnetic fields, rotation and chemical 
abundances evolve over time is not well understood.  
Temporal changes in rotation speed and abundance patterns potentially can provide valuable insights into the underlying 
mechanisms producing the abundance anomalies and magnetic fields seen in Ap stars. 
In that context, this paper serves as a first step in a larger effort to examine trends in the evolution of Ap stars.  
The use of cluster stars allows us, in principle, to remove the competing effects of age and environment, and allows 
for precise determination of mass and evolutionary state, particularly for stars in the earlier half of their lifetime.  
Through the use of a number of clusters with different ages, one can construct a picture of evolutionary trends.  
Thus the detailed chemical abundances of all Ap stars, within a single cluster, presented in this study provides 
important groundwork for a description of the formation and evolution of abundances of Ap stars.  

\section*{Acknowledgments} We would like to thank ESO and the POP for collecting the spectra used in this project and making them publicly available.  
This research has made use of the WEBDA database, developed by J.-C. Mermilliod and 
maintained by E. Paunzen at the Institute for Astronomy of the University of Vienna. CPF and GAW acknowledge 
support from the Academic Research Programme of the Canadian Department of National Defence. CPF, GAW and JDL 
acknowledge Discovery Grant support from the Natural Sciences and Engineering Research Council of Canada.

{}

\label{lastpage}

\end{document}